\documentclass[12pt]{article}
\usepackage{graphicx}
\usepackage{amsbsy}
\usepackage{amstext}
\usepackage{amsmath}
\usepackage{amssymb}
\usepackage{cite}

\newcommand\pubdate{\today}

\def\IPNL{Universit\'{e} Lyon 1 \& CNRS/IN2P3, UMR5822 IPNL\\
4 rue Enrico Fermi, 69622 Villeurbanne Cedex, FRANCE}

\def\Title#1{\begin{center} {\Large #1 } \end{center}}
\def\Author#1{\begin{center}{ \sc #1} \end{center}}
\def\Address#1{\begin{center}{ \it #1} \end{center}}

\newcommand\pubblock{\rightline{\begin{tabular}{l} \pubdate  \end{tabular}}}
\newenvironment{Abstract}{\begin{quotation}  }{\end{quotation}}
\newenvironment{Presented}{\begin{quotation} \begin{center} 
             PRESENTED AT\end{center}\bigskip 
      \begin{center}\begin{large}}{\end{large}\end{center} \end{quotation}}
\def\Acknowledgements{\bigskip  \bigskip \begin{center} \begin{large}
             \bf ACKNOWLEDGEMENTS \end{large}\end{center}}

\def\beq{\begin{equation}}
\def\eeq#1{\label{#1}\end{equation}}
\def\eeqn{\end{equation}}

\def\beqa{\begin{eqnarray}}
\def\eeqa#1{\label{#1}\end{eqnarray}}
\def\eeqan{\end{eqnarray}}

\let\bar=\overbar

\def\Dslash{\not{\hbox{\kern-4pt $D$}}}
\def\dslash{\not{\hbox{\kern-2pt $\del$}}}

\def\msb{{\bar{\ssstyle M \kern -1pt S}}}

\begin{document}
\begin{titlepage}
\pubblock

\vfill
\Title{Rare K and B decays with non-standard missing energy}
\vfill
\Author{ Christopher Smith\footnote{c.smith@ipnl.in2p3.fr}}
\Address{\IPNL}
\vfill
\begin{Abstract}
The rare K and B semileptonic decays into neutrino pairs are well-known to be extremely sensitive to non-standard physics in the quark sector. In this talk, their capabilities to signal New Physics in the leptonic sector, or even to reveal entirely new invisible sectors, are analyzed.
\end{Abstract}
\vfill
\begin{Presented}
CKM2010\\6th International Workshop on the CKM Unitarity Triangle\\University of Warwick, UK, 6-10 September 2010
\end{Presented}
\vfill
\end{titlepage}
\def\thefootnote{\fnsymbol{footnote}}
\setcounter{footnote}{0}

\section{Introduction}

The rare $P\rightarrow P^{\prime }\nu\bar{\nu}$ decays with $P=K,B$ are fantastic windows into the flavor-changing neutral current (FCNC) transitions $s\rightarrow d$ and $b\rightarrow s,d$. Indeed, being induced at loop level~\cite{SMB,SMK,Matrix,Tau}, and only through GIM breaking effects, they are significantly suppressed, allowing New Physics (NP) to easily be competitive. In addition, the Z penguin driving these FCNC is effectively a dimension-four operator after the electroweak symmetry breaking, and thus especially sensitive to the physics at the electroweak scale. 
The NP impacts have been extensively studied. Let us just mention (see e.g. Ref.~\cite{NPB}) the MSSM, little Higgs model, extra dimensions, fourth generation, unparticles... 
In the present talk, we will look at the rare $P\rightarrow P^{\prime }\nu\bar{\nu}$ decays from a different perspective. Since the neutrinos are not seen, these decays are experimentally undistinguishable from $P\rightarrow P^{\prime }+X$ with $X$ any neutral invisible state. So, we would like to review briefly a few scenarios where $X\neq \nu_{e,\mu,\tau}\bar{\nu}_{e,\mu,\tau}$.

\begin{figure}[tb]
\centering\includegraphics[width=0.93\textwidth]{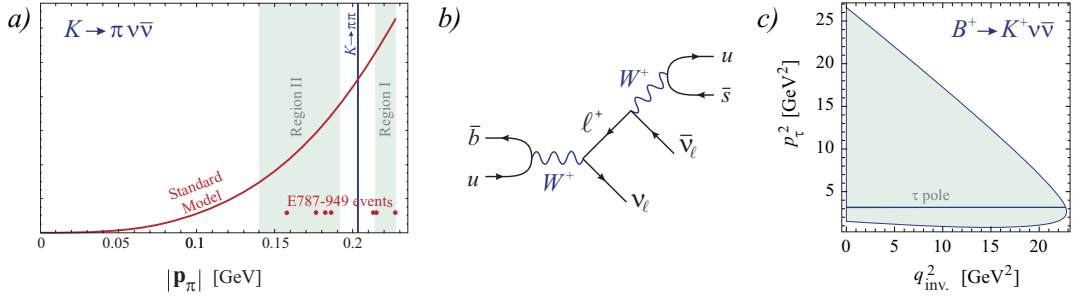}
\caption{ {\small $a)$ The $K^{+}\rightarrow \pi ^{+}\nu \bar{\nu}$ differential rate in the SM, with the NA62 windows and E787-949 events. $b)$ Tree-level process for $P\rightarrow P^{\prime }\nu \bar{\nu}$. $c)$ Dalitz plot for $B^{+}\rightarrow K^{+}\nu \bar{\nu}$ ($B^{+}\rightarrow (\pi^{+},K^{\ast+})\nu \bar{\nu}$ are similar).}}
\label{Fig1}
\end{figure}

\vskip 5pt
\noindent\textbf{Kinematics and observables.} To set the stage, let us first consider $P\rightarrow P^{\prime }\nu\bar{\nu}$ from an experimental perspective. In the $P$ rest frame, all that is seen (or reconstructed) is the momentum of $P^{\prime }$, directly related to the $\nu\bar{\nu}$ invariant mass $z=(p_{\nu }+p_{\bar{\nu}})^{2}/m_{K}^{2}$ (``missing energy''). So, only the differential rate $\partial \Gamma /\partial z$ is accessible, and given the very small rates, only in those momentum regions where backgrounds can be controlled. The total rate is then extrapolated assuming the SM shape for $\partial \Gamma/\partial z$.

For $K\rightarrow \pi \nu \bar{\nu}$ in the SM, the dominant contributions from the $Z$ penguin and $W$ boxes with $t$, $c$ quarks is encoded in an effective $\left( \bar{s}d\right) _{V-A}\otimes \left( \bar{\nu}\nu \right)_{V-A}$ interaction. With the relevant form-factors of the $\langle \pi |\left( \bar{s}d\right)_{V}|K\rangle $ matrix elements extracted with per-mil precision from $K_{\ell 3}$ decays~\cite{Matrix}, the SM differential rate is completely fixed, see Fig.~\ref{Fig1}.

For the $B\rightarrow (\pi,K^{(\ast )})\nu \bar{\nu}$ decays in the SM, the situation is similar except for $B^{+}$, for which the tree-level $B^{+}\rightarrow \nu _{\tau }[\tau^{+}\rightarrow (\pi ,K^{(\ast )})^{+}\bar{\nu}_{\tau }]$ mechanism opens up~\cite{Tau}. Actually, this process could be considered as $B^{+}\rightarrow \nu _{\tau }\tau ^{+}$, with the $\tau $ reconstructed using its hadronic decays. However, it is not possible to kinematically disentangle the tree-level and loop-induced FCNC processes because the $\tau $ pole runs over most of the missing energy range, see Fig.~\ref{Fig1}. The consequences are different for $\Delta S=0$ and $\Delta S=1$ decays because the $Z$ penguin scales like $V_{tb}^{\dagger }V_{ts/d}$, and the $\tau $ pole contribution like $V_{ub}^{\dagger }V_{us/d}$. So, taking the $B^{+}\rightarrow \nu _{\tau }\tau ^{+}$ perspective, (1) $B^{+}\rightarrow \nu _{\tau }[\tau ^{+}\rightarrow leptonic]$ is safe, since there is clearly no FCNC pollution. (2) $B^{+}\rightarrow \nu_{\tau }[\tau ^{+}\rightarrow \pi ^{+}\bar{\nu}_{\tau }]$ is dominated by the charged current, with only a 2\% pollution from the $b\rightarrow d\nu\bar{\nu}$ FCNC. So, NP in $b\rightarrow d\nu \bar{\nu}$ can only be probed with $B^{0}$ decays. (3) $B^{+}\rightarrow \nu _{\tau }[\tau ^{+}\rightarrow K^{(\ast )+}\bar{\nu}_{\tau }]$ is apparently enhanced by 600\%-700\% due to the large $b\rightarrow s\nu \bar{\nu}$ FCNC contribution, and thus sensitive to NP both in $B^{+}\rightarrow \nu _{\tau }\tau ^{+}$ and in $b\rightarrow s\nu \bar{\nu}$.

\section{Lepton flavor violating effects}

Since the neutrino flavors are not detected, experiments actually probe $P\rightarrow P^{\prime }\nu ^{I}\bar{\nu}^{J}$ with both $I=J$ and $I\neq J$. However, if we assume that NP respects the SM gauge invariance, then the left-handed neutrinos are together with charged leptons in weak doublets, and $P\rightarrow P^{\prime }\nu ^{I}\bar{\nu}^{J}$ are necessarily correlated to $P\rightarrow P^{\prime }\ell ^{I}\bar{\ell}\hspace{0in}^{J}$ for which flavor information is readily accessible. So, given the current experimental bounds on these latter modes, we can get an idea of the size of $P\rightarrow P^{\prime }\nu ^{I}\bar{\nu}^{J}$, $I\neq J$ processes.

\vskip 5pt
\noindent\textbf{The flavor puzzle.} To be more specific, writing the SM matter fields as $Q^{T}=(u_{L},d_{L})$, $U=\bar{u}_{R}$, $D=\bar{d}_{R}$, $L^{T}=(\nu _{L},\ell_{L})$, and $E=\bar{\ell}_{R}$, the generic four-fermion, dimension-six effective operators are $\mathcal{H}_{eff}=\Lambda ^{-2} C_{i}^{IJKL}O_{i}^{IJKL}$, $O_{1}=\bar{Q}\Gamma Q\otimes \bar{L}\Gamma L$, $O_{2}=D\Gamma \bar{D}\otimes \bar{L}\Gamma L$, $O_{3}=D\Gamma \bar{D}\otimes E\Gamma \bar{E}$, $O_{4}=\bar{Q}\Gamma Q\otimes E\Gamma \bar{E}$, $O_{5}=\bar{Q}\Gamma \bar{D}\otimes E\Gamma L$, where the quark and lepton flavor indices $I,J,K,L$ are understood and $\Gamma$ stands for all possible Dirac and $SU(2)_L$ structures. Only $O_{1,2}$ contribute to $P\rightarrow P^{\prime }\nu ^{I}\bar{\nu}^{J}$, but all of them induce $P\rightarrow P^{\prime }\ell ^{I}\bar{\ell}\hspace{0in}^{J}$. Typically, given the general agreement of observed $P\rightarrow P^{\prime}\ell ^{I}\bar{\ell}\hspace{0in}^{I}$ and $P\rightarrow P^{\prime }\nu ^{I}\bar{\nu}^{I}$ rates with the SM predictions, as well as the tight bounds on $P\rightarrow P^{\prime }\ell^{I}\bar{\ell}\hspace{0in}^{J}$, $I\neq J$, generic $C_{i}^{IJKL}\sim \mathcal{O}(1)$ couplings are not compatible with a relatively low NP scale, $\Lambda \lesssim 1$ TeV. This is the NP flavor puzzle~\cite{CERN}.

\vskip 5pt
\noindent\textbf{What Minimal Flavor Violation can say?} If TeV-scale NP is present, its flavor structures must be highly non trivial, but without any further input, no specific conclusions could be drawn about the sizes of the $C_{i}^{IJKL}$. So, to proceed, we will consider the specific theoretical setting called Minimal Flavor Violation~\cite{MFV} where the NP flavor structures are compelled to be aligned with those of the SM.

To do this, a well-define procedure relies on the global $SU(3)^{5}$ symmetry exhibited by the SM (or MSSM) gauge interactions, arising from their complete flavor blindness. In the SM, this symmetry is only broken by the Yukawa couplings, but in presence of NP, the extended flavor sector would also break the $SU(3)^{5}$ symmetry. The MFV hypothesis limits the set of possible $SU(3)^{5}$ breaking terms to just that of the SM, and further promotes these terms to $SU(3)^{5}$ spurions. Crucially, the NP flavor couplings then become expressible entirely out of the spurions, i.e. the Yukawa couplings, and thus naturally inherit their very peculiar hierarchies. Allowing for neutrino masses using a Type I seesaw, the minimal spurion content needed to account for fermion masses is $v\mathbf{Y}_{u}=\mathbf{m}_{u}V$, $v\mathbf{Y}_{d}=\mathbf{m}_{d}$, $v\mathbf{Y}_{e}=\mathbf{m}_{e}$, and $v^{2}\mathbf{Y}_{\nu }^{T}(\mathbf{M}_{R}^{-1})\mathbf{Y}_{\nu }=U^{\ast }\mathbf{m}_{\nu }U^{\dagger }$, where $\mathbf{m}_{u,d,e,\nu }$ are diagonal, $\mathbf{M}_{R}$ is the heavy $\nu_{R}$ mass matrix, and $v$ the Higgs vacuum expectation value. The CKM matrix $V$ is put in $\mathbf{Y}_{u}$ so that the down quarks are mass eigenstates, and similarly for the PMNS matrix $U$. In most models, there is yet another unsuppressed spurion induced by the seesaw mechanism, $\mathbf{Y}_{\nu}^{\dagger }\mathbf{Y}_{\nu }$, though it cannot be entirely fixed~\cite{CasasI01} from $\mathbf{m}_{\nu }$ and $U$, even assuming $\mathbf{M}_{R}=M_{R}\mathbf{1}$.

Returning to our operators, and first enforcing MFV on their quark structures $\bar{Q}Q$, $D\bar{D}$ or $DQ$, we see that only $\bar{Q}Q$ is unsuppressed. Indeed, we need to construct an $SU(3)^{5}$ symmetric combination of Yukawa couplings involving the non-diagonal $\mathbf{Y}_{u}^{\dagger }\mathbf{Y}_{u}$ to allow for the $s\rightarrow d$, $b\rightarrow d$ or $b\rightarrow s$ flavor transitions. But for external $D$ quarks, this means inserting first the diagonal $\mathbf{Y}_{d}$, i.e. light quark masses. The same holds for the lepton side, so the dominant LFV effects should be induced by $\bar{Q}\hspace{0in}^{I}(\mathbf{Y}_{u}^{\dagger }\mathbf{Y}_{u})^{IJ}Q^{J}\otimes \bar{L}\hspace{0in}^{K}(\mathbf{Y}_{\nu }^{\dagger }\mathbf{Y}_{\nu})^{KL}L^{L}$, for which $\mathcal{B}(P\rightarrow P^{\prime }\nu ^{I}\bar{\nu}^{J})\sim \mathcal{B}(P\rightarrow P^{\prime }\ell^{I}\bar{\ell}\hspace{0in}^{J})$.

\vskip 5pt
\noindent\textbf{Numerical estimates.} The $\mathbf{Y}_{u}^{\dagger }\mathbf{Y}_{u}$ insertion naturally induce the SM scalings
\begin{equation}
s\rightarrow d\sim |V_{td}V_{ts}^{\dagger }|\sim \lambda^{5},\;\;b\rightarrow d\sim |V_{td}V_{tb}^{\dagger }|\sim \lambda^{3},\;\;b\rightarrow s\sim |V_{ts}V_{tb}^{\dagger }|\sim \lambda ^{2}\;.
\label{SMCKM}
\end{equation}%
The NP contributions to both $P\rightarrow P^{\prime }\nu ^{I}\bar{\nu}^{I}$ and $P\rightarrow P^{\prime }\ell ^{I}\bar{\ell}\hspace{0in}^{I}$ thus cannot exceed the SM ones, even with $\Lambda \lesssim 1$ TeV, and the NP flavor puzzles are solved. Turning to the LFV part, it necessarily scales like the off-diagonal entries of $\mathbf{Y}_{\nu }^{\dagger }\mathbf{Y}_{\nu }$. Since $\mathbf{m}_{\nu }\sim \mathbf{Y}_{\nu }^{T}(\mathbf{M}_{R}^{-1})\mathbf{Y}_{\nu }$, it would appear that taking $\mathbf{M}_{R}$ sufficiently large would ensure $\mathbf{Y}_{\nu }\sim \mathcal{O}(1)$, leading to $P\rightarrow P^{\prime }\nu ^{I}\bar{\nu}^{J}\sim P\rightarrow P^{\prime }\nu ^{I}\bar{\nu}^{I}$. However, such large $\mathbf{Y}_{\nu }$ are forbidden by $\ell^{I}\rightarrow \ell ^{J}\gamma$, tuned by $E\mathbf{Y}_{e}(\mathbf{Y}_{\nu }^{\dagger }\mathbf{Y}_{\nu})\sigma ^{\mu \nu }LH^{\dagger }F_{\mu \nu }$. Conservatively, we can at most get $(\mathbf{Y}_{\nu }^{\dagger }\mathbf{Y}_{\nu })^{I\neq J}\sim 1\%$, so that for $I\neq J$, $L=\nu,\ell$, $\mathcal{B}^{MFV}(K\rightarrow\pi L^{I}\bar{L}\hspace{0in}^{J})\lesssim 10^{-15}$ and $\mathcal{B}^{MFV}(B\rightarrow (\pi ,K)L^{I}\bar{L}\hspace{0in}^{J})\lesssim 10^{-10}$, well beyond experimental reach.

It is worth to trace the origin of this suppression. Within MFV, the quark and lepton flavor sectors are completely disconnected since the flavor group factorizes as $SU(3)^5=SU(3)^3_q \otimes SU(3)^2_{\ell}$. Then, the quark transitions are constrained from lepton flavor conserving processes, while the LFV transitions are bounded from purely leptonic observables. By contrast, in a model where LFV occurs only together with a quark transition, as for example through leptoquark exchanges~\cite{LQ}, the $\ell ^{I}\rightarrow \ell ^{J}\gamma $ constraints would be alleviated and larger effects may be possible. So, the search for these modes not only tests the SM, but also MFV in its most general implementation.

\section{R-parity violating effects}

In the MSSM, baryon ($\mathcal{B}$) and lepton ($\mathcal{L}$) numbers are not automatically conserved, because squarks and sleptons are $\mathcal{B}$ and $\mathcal{L}$ carrying scalars. So, the superpotential can contain the renormalizable terms $\boldsymbol{\mu}^{\prime }LH_{d}+\boldsymbol{\lambda}LLE+\boldsymbol{\lambda}^{\prime }LQD+\boldsymbol{\lambda}^{\prime \prime}UDD$. However, the sfermions then induce new dilepton, diquark or leptoquark currents, which can drive proton decay. Experimentally, $\tau_{p^{+}}\gtrsim 10^{31}$ years~\cite{PDG}, so these couplings must be so small that one usually forbids them by imposing a new symmetry, R-parity.

\begin{figure}[t]
\centering\includegraphics[width=0.98\textwidth]{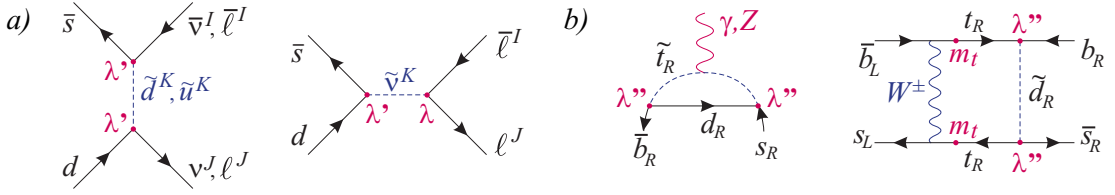}
\caption{ {\small $a)$ Tree-level contributions to $P\rightarrow P^{\prime }\protect\nu ^{I}\bar{\protect\nu}^{J}$ and $P\rightarrow P^{\prime }\ell ^{I}\bar{\ell}\hspace{0in}^{J}$ decays, induced by $\Delta \mathcal{L}$ couplings. $b)$ Loop-level FCNC induced by $\Delta \mathcal{B}$ couplings.}}
\label{Fig2}
\end{figure}

\vskip 5pt
\noindent\textbf{Possible effects on rare decays.} To escape the $\tau_{p^{+}}$ bounds, but nevertheless have signals in semileptonic decays, the usual strategy is to impose only the conservation of $\mathcal{B}$. Then, the rare decays arise at tree-level~\cite{RPV}, see Fig.~\ref{Fig2}. After Fierzing the leptoquark currents, the possible operators are $\boldsymbol{\lambda}^{\prime 2}\bar{Q}\gamma_{\mu }Q\otimes L\gamma^{\mu }L$, $\boldsymbol{\lambda}^{\prime 2}D\gamma_{\mu }\bar{D}\otimes L\gamma^{\mu }L$, and $\boldsymbol{\lambda}^{\prime}\boldsymbol{\lambda}\bar{Q}\;\bar{D}\otimes EL$. So, this scenario collapses to that discussed in the previous section, with $P\rightarrow P^{\prime }\nu ^{I}\bar{\nu}^{J}$ correlated with $P\rightarrow P^{\prime}\ell ^{I}\bar{\ell}\hspace{0in}^{J}$, except that it apparently evades the model-independent MFV constraints. So, let us now look more closely at how MFV can be enforced for R-parity violating couplings. 

\vskip 5pt
\noindent\textbf{What Minimal Flavor Violation can say?} Since the $\Delta \mathcal{B}$ and $\Delta\mathcal{L}$ interactions are flavored, the very long proton lifetime could be explained by the same mechanism as the other flavor puzzles. So, instead of imposing R-parity, let us extend the MFV framework to $\Delta \mathcal{B}$ and $\Delta\mathcal{L}$ interactions, i.e. force them to be entirely constructed out of the SM spurions~\cite{MFVRPV}. The crucial observation, allowing MFV to be a viable alternative, is the different $SU(3)^{5}$ symmetry properties of the $\Delta \mathcal{B}$ and $\Delta \mathcal{L}$ couplings. Indeed, while the $\boldsymbol{\lambda}^{\prime \prime }$ couplings can be parametrized in terms of $\mathbf{Y}_{u,d}$, the $\boldsymbol{\mu}^{\prime }$, $\boldsymbol{\lambda}$, and $\boldsymbol{\lambda}^{\prime }$ couplings require a spurion transforming like $\boldsymbol{6}\sim v^{2}\mathbf{Y}_{\nu }^{T}(\mathbf{M}_{R}^{-1})\mathbf{Y}_{\nu }=U^{\ast }\mathbf{m}_{\nu }U^{\dagger }$. So, the $\Delta \mathcal{L}$ couplings are tuned by the tiny $\nu_L$ masses. In addition, $\Delta \mathcal{B}$ and $\Delta \mathcal{L}$ couplings require the antisymmetric tensors of $SU(3)^{5}$ to form invariants, bringing in small fermion mass factors. Altogether, $\Delta\mathcal{B}$ and $\Delta\mathcal{L}$ couplings are sufficiently suppressed to pass all their experimental bounds~\cite{MFVRPV}.

Returning to the rare decays, the only possible mechanisms within MFV involve exclusively the $\boldsymbol{\lambda}^{\prime \prime }$ couplings, since the others are negligible. The loop-level FCNC processes (Fig.~\ref{Fig2}) then produce SM-like $\nu^I\bar{\nu}^I$ final states, and scale like~\cite{MFVRPV}
\[
s\rightarrow d\sim |\boldsymbol{\lambda}_{323}^{\prime \prime \ast }\boldsymbol{\lambda}_{331}^{\prime \prime }|<10^{-8},b\rightarrow d\sim | \boldsymbol{\lambda}_{312}^{\prime \prime \ast }\boldsymbol{\lambda}_{323}^{\prime \prime }|<10^{-5},b\rightarrow s\sim |\boldsymbol{\lambda}_{312}^{\prime \prime \ast }\boldsymbol{\lambda}_{331}^{\prime \prime}|<10^{-3}.
\]
Compared to the SM scalings of Eq.~(\ref{SMCKM}), the only place where one could hope to see an effect is for $b\rightarrow s$ transition, though not beyond the few percent level.

\section{Very light invisible particles}

Experimentally, a measurement of $P\rightarrow P^{\prime }\nu \bar{\nu}$ would implicitly includes any other invisible final states $P\rightarrow P^{\prime }X$, where $X$ denotes either a single state or a collection of particles with some unknown spin and mass. To ensure its invisibility, let us assume that $X$ is made of scalars under all the SM gauge interactions. Then, the main question is whether the flavor-changing rare decays could compete with flavor-blind searches like for example the EWPO or quarkonium decay.

\begin{figure}[t]
\centering\includegraphics[width=0.98\textwidth]{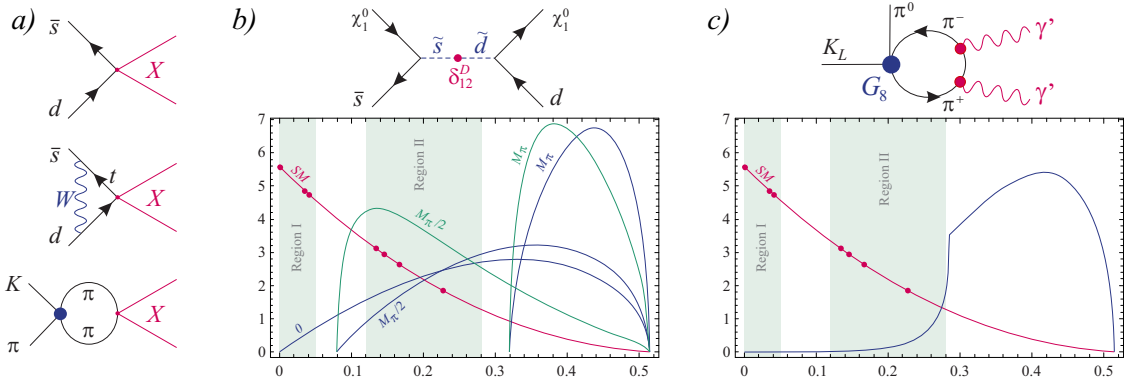}
\caption{ {\small $a)$ Flavor-based classification of the couplings to invisible states. $b)$ Light neutralino as an example of flavor-changing invisible particles. Green (blue) is for the axial-vector (scalar) effective couplings. $c)$ Extra photon as an example of flavor-blind invisible particles coupled to light quarks.}}
\label{Fig3}
\end{figure}

We can distinguish several scenarios (see Fig.~\ref{Fig3}). For couplings of the form $d^{I}\Gamma d^{J}X$ with $I\neq J$, the production of $X$ suffices to induce the weak transitions. Given their smallness in the SM, rare decays would not only be the natural observables to search for such invisible states, they would also be very sensitive. On the contrary, if $X$ has only flavor-blind couplings, one has to rely on the weak interactions to induce the flavor transition, at the cost of a $G_{F}^{2}$ factor for the rate. Rare decays may still be competitive if $X$ couples only to heavy quarks, thereby inducing new FCNC, essentially because $t\bar{t}X$ is not easy to constrain otherwise. For flavor-blind couplings to light quarks, the situation is less clear. Naively, not least because of their inherently long-distance nature, rare decays appear superseded by flavor blind observables, though possibly not over the whole $X$ mass range. Nevertheless, as we will see, there may be special situations where competitive bounds could be obtained.

A complete study is in progress~\cite{Invs}, so here let us simply present two examples sitting at opposite ends of the spectrum of possible $X$ couplings.

\vskip 5pt
\noindent\textbf{Example 1: Very light neutralinos.} As an example of flavor-changing couplings of $X$, consider the situation where the lightest (stable) neutralino has a mass in the MeV range. Relying on either the soft SUSY-breaking squark masses or the trilinear terms to induce the flavor transitions, down squark tree-level exchanges can lead to the effective operator $(\bar{s}\gamma ^{\mu }(1\pm \gamma _{5})d)\otimes (\bar{\chi}\gamma _{\mu }\gamma _{5}\chi )$ or $(\bar{s}(1\pm \gamma_{5})d)\otimes (\bar{\chi}(1\pm \gamma _{5})\chi )$, respectively (and similarly for $B$ decays). As shown in Fig.3, for sufficiently light $\chi _{1}^{0}$, the experimental regions overlap well with the signals from $K\rightarrow \pi\chi _{1}^{0}\chi _{1}^{0}$, especially for the axial vector effective coupling. See Ref.~\cite{LChi} for more information.

\vskip 5pt
\noindent\textbf{Example 2: Weakly coupled extra photon.} As an example of a flavor blind scenario with couplings only to light quarks, imagine that there is an extra massless vector with couplings aligned with those of the photon~\cite{PhotP}. Let us specialize to the $K$ decay case, for which the interactions $e^{\prime }A_{\mu }^{\prime }\sum_{q=u,d,s}Q_{q}\bar{q}\gamma^{\mu }q$ are easily incorporated within Chiral Perturbation Theory. The dominant decay mechanism is shown in Fig.~\ref{Fig3}. Actually, a precise computation is not even needed since obviously $\mathcal{B}(K_{L}\rightarrow \pi^{0}\gamma^{\prime }\gamma^{\prime})\approx \alpha^{\prime 2}/\alpha^{2}\mathcal{B}(K_{L}\rightarrow\pi ^{0}\gamma \gamma)$, with $\alpha ^{\prime }=e^{\prime 2}/4\pi$. This perfectly illustrates the high cost of the weak transition: $\mathcal{B}(K_{L}\rightarrow \pi^{0}\gamma \gamma )$ being already in the $10^{-6}$ range, and even with about $10^{13}$ kaons so as to access $\mathcal{B}(K_{L}\rightarrow \pi^{0}\nu \bar{\nu})\sim 10^{-11}$, only the region where $\alpha ^{\prime }/\alpha \gtrsim 10^{-3}$ would be probed. This is excluded by flavor-blind experiment, at least if $\alpha_q\neq\alpha_{\ell}$ holds. More problematically for rare K experiments, $K_{L}\rightarrow \pi ^{0}\gamma ^{\prime }\gamma ^{\prime }$ is only sizeable above the $K \rightarrow 3 \pi$ threshold, see Fig.~\ref{Fig3}.

For this scenario, it may be much more interesting to look for $K_{L}\rightarrow \pi ^{0}\gamma \gamma ^{\prime }$, since then one may theoretically reach down to $\alpha ^{\prime }/\alpha \approx 10^{-7}$. So, provided $K\rightarrow \pi \gamma $ + (missing energy) is experimentally accessible for sufficiently high $\gamma\gamma^{\prime}$ invariant masses, the future $K$ experiments would be competitive with flavor-blind searches~\cite{Invs}.

\section{Conclusion}

The rare K and B decays are well-known to be extremely sensitive to non-standard physics inducing the quark transitions, to the point that even in the LHC era, they are our best windows into possible extensions of the flavor sector. In this talk, the capabilities of these decays to signal New Physics in the leptonic sector, or even to reveal entirely new invisible sectors, have been analyzed. Given their unique and so broad sensitivities, these decays will most certainly play a central role in the future.

\Acknowledgements
Many thanks to the convenors of the WG III - Rare decays for their kind invitation.

\end{document}